\documentstyle[11pt]{jcp3b}
\def\PLOTDIR{.}
\input{psfig}

\newcommand{\beq}{\begin{equation}}
\newcommand{\eeq}{\end{equation}}
\newcommand{\beqa}{\begin{eqnarray}}
\newcommand{\eeqa}{\end{eqnarray}}
\newcommand{\half}{\frac{1}{2}}

\newcommand{\r}{{\bf r}}

\newcommand{\ri}{\r_i}

\newcommand{\rj}{\r_j}

\newcommand{\Gij}{G_{ij}}

\newcommand{\Ginv}{G^{-1}}
\newcommand{\Ginvij}{\Ginv_{ij}}

\newcommand{\sqtopi}{\sqrt{\frac{2}{\pi}}}

\newcommand{\la}{\langle}
\newcommand{\ra}{\rangle}
\renewcommand{\(}{\left(}
\renewcommand{\)}{\right)}

\begin{document}
\begin{flushright}
LU TP 96-30\\
December 23, 1996\\
\end{flushright}
\vspace{0.5in}
\LARGE
\begin{center}
{\bf The Electrostatic Persistence Length Calculated from Monte Carlo,
Variational and Perturbation Methods}\\
\vspace{.3in}
\large
Magnus Ullner\footnote{fk2mul@dix.fkem2.lth.se} and
Bo J\"{o}nsson\footnote{fk2boj@grosz.fkem2.lth.se}\\
\vspace{0.05in}
Physical Chemistry 2, Chemical Center, University of Lund\\ Box 124, S-221 00 Lund, Sweden\\
\vspace{0.15in}
Carsten Peterson\footnote{carsten@thep.lu.se},
Ola Sommelius\footnote{ola@thep.lu.se} and Bo
S\"{o}derberg\footnote{bs@thep.lu.se}\\
\vspace{0.05in}
Complex Systems Group, 
Department of Theoretical Physics\\ University of Lund, S\"{o}lvegatan 14A,
S-223 62 Lund, Sweden\\
\vspace{0.15in}

Submitted to {\it Journal of Chemical Physics}

\end{center}
\normalsize

Abstract:

Monte Carlo simulations and variational calculations using a Gaussian ansatz
are applied to a model consisting of a flexible linear polyelectrolyte
chain as well as to an intrinsically stiff chain with
up to 1000 charged monomers. Addition of salt is treated implicitly
through a screened Coulomb potential for the electrostatic interactions.

For the flexible model 
the electrostatic persistence length shows roughly three regimes in its
dependence on the Debye-H\"{u}ckel screening length, $\kappa^{-1}$.
As long as the salt content is low and $\kappa^{-1}$ is longer
than the end-to-end distance, the electrostatic persistence length varies
only slowly with $\kappa^{-1}$. Decreasing the screening length, a controversial
region is entered. We find that the electrostatic persistence length scales
as $\sqrt{\xi_p}/\kappa$, in agreement with experiment on flexible polyelectrolytes,
where $\xi_p$ is a strength
parameter measuring the electrostatic interactions within the polyelectrolyte.
For screening lengths much shorter than the bond length, the $\kappa^{-1}$
dependence becomes quadratic in the variational calculation. The simulations
suffer from numerical problems in this regime, but seem to
give a relationship half-way between linear and quadratic.

A low temperature expansion only reproduces the first regime and a high
temperature expansion, which treats the electrostatic interactions as a
perturbation to a Gaussian chain, gives a quadratic dependence on the Debye
length.

For a sufficiently stiff chain, the persistence length varies quadratically
with $\kappa^{-1}$ in agreement with earlier theories.

\normalsize

\newpage

\section{INTRODUCTION}
During the last decade or so a large number of simulations of charged
polymers have appeared in the literature. Many of these have been devoted to the
study of conformational properties in bulk
solution,\cite{Valleau:89,Hooper:90,Reed:91a,Christos:92,Stevens:94,Seidel:96}
while a few have focused on the
dramatic effects polyelectrolytes can produce in charged colloidal
dispersions.\cite{Akesson:89,Wallin:96} In
the former case the polymer chain is quite often characterized by a few geometric
parameters like the end-to-end separation or the radius of gyration. The behaviour of
a polyelectrolyte has also been analyzed in terms of its persistence length.
The subject of persistence length may sometimes appear confusing due to
the number of definitions and different ways to calculate the quantity.
The general concept of persistence
length is however the same --- it is a measure of stiffness, a characteristic
length over which a polymer chain remembers its direction.
Of course, even a neutral polymer will have a non-zero persistence length
and it has become common practice to divide the total persistence length into an
electrostatic, $l_{p,e}$, and a non-electrostatic or "bare" part, $l_{p,0}$,
\begin{equation}
l_{p,t} = l_{p,0} + l_{p,e} .
\end{equation}
Odijk\cite{Odijk:77} and, independently, Skolnick and Fixman\cite{Skolnick:77}
(OSF) derived equivalent models to calculate the influence of electrostatic
interactions on the persistence length and in particular the effect of screening
from added salt. For some time there has been a controversy in the literature
over conflicting results for the ionic strength dependence
\cite{Barrat:93a,Ha:95,Micka:96,Micka:97} and it is
the purpose of this communication to resolve the conflict. 

Experimentally the OSF result, which suggests a quadratic dependence on the
Debye length, $\kappa^{-1}$, for the electrostatic persistence length, has been confirmed for
intrinsically very stiff polyelectrolytes, like DNA\cite{Maret:83} and
poly(xylylene tetrahydrothiophenium chloride)\cite{Mattoussi:92}, but
although it has been claimed that the theory is also valid for long, flexible
chains,\cite{Odijk:78,Khokhlov:82} the experimental findings generally points
towards a linear relationship, i.e. $l_{p,e}\sim\kappa^{-1}$, for flexible
polyelectrolytes.\cite{Tricot:84,Ghosh:90,Reed:91b,Degiorgio:91} An objection to the OSF 
theory is the neglect of entropy. Taking this
into account, variational calculations also give the linear
relationship.\cite{Schmidt:91,Barrat:93a,Ha:95} Simulations usually
show deviations from the OSF theory, and the exponent $p$ in $l_{p,e}\sim
\kappa^{-p}$ may vary from 1.5\cite{Seidel:94} down to
0.3-0.9\cite{Micka:96}. However, $p \approx 1$, in agreement with experiment,
have also been reported.\cite{Reed:91a,Seidel:96} 

It is our conclusion that
an almost linear relationship does exist for a flexible polyelectrolyte when
the screening length is shorter than the dimensions of the
chain, but longer than the monomer-monomer separation. 
For an intrinsically stiff chain, however, we find a quadratic dependence in
agreement with the OSF theory.

\section{THE PERSISTENCE LENGTH}
Prior to discussing the models and our different approaches we review different
definitions of the persistence length and briefly discuss the OSF theory.
Given a linear chain with $N$ monomers, the microscopic approach to the
persistence length is simply to
sum the average 
scalar products of bond vectors with respect to a certain bond, i.e. to
calculate the average sum of directional cosines,
\begin{equation}
l_{p,i}=\frac{1}{a} \sum_{k=0}^{N-1-i}\langle {\bf r}_i\cdot{\bf r}_{i+k} \rangle =
a \sum_{k=0}^{N-1-i}\langle \cos \theta_{i+k} \rangle .
\label{elpi}
\end{equation}
where ${\bf r}_i={\bf x}_{i+1} - {\bf x}_i$ is the bond vector between monomer $i$ and
$i+1$, whereas ${\bf x}_i$ represents the coordinate of monomer $i$ and $a$ is an
average bond length. In the following $a$ will represent the root-mean-square
monomer-monomer separation,
\begin{equation}
a^2=\frac{1}{N-1} \sum_{j=1}^{N-1}\langle {\bf r}_j^2 \rangle,
\end{equation}
which, of course, is the same as the bond length for a chain with rigid bonds.
Usually, $i$ denotes the central bond
and the cosines are calculated from the average of the scalar products in both
directions
and not just for subsequent bonds as written in Eq.~(\ref{elpi}). One may also
calculated $l_{p,N/2}$ as an average over both directions. The average
over all the local and uni-directional persistence lengths, $l_{p,i}$,
is related to the mean-square end-to-end separation,
\begin{equation}
R_{ee}^2\equiv \langle \left( {\bf x}_N-{\bf x}_1 \right)^2 \rangle=
\langle \sum_{i=1}^{N-1} {\bf r}_i \cdot \sum_{j=1}^{N-1} {\bf r}_j \rangle =
\sum_{i=1}^{N-1} \langle {\bf r}_i^2 \rangle + 2 \sum_{i=1}^{N-2} \sum_{j=i+1}^{N-1} 
\langle {\bf r}_i \cdot {\bf r}_j \rangle,
\label{eRee}
\end{equation}
i.e.
\begin{equation}
l_{p,R}\equiv \frac{1}{N-1} \sum_{i=1}^{N-1} l_{p,i} = \frac{R_{ee}^2}{2(N-1)a}
+\frac{a}{2}.
\label{elpR}
\end{equation}

In the case of an infinite chain, all bonds are equivalent and so are
the persistence lengths of
Eqs. (\ref{elpi}) and (\ref{elpR}). We may in fact treat the polymer as infinite
and disregard the index $i$ in Eq. (\ref{elpi}), if the chain is long enough
to make end-effects unimportant, i.e., for most bonds $\langle \cos \theta_{i+k} \rangle$
will become zero before $k$ reaches the end and most of the chain
will be described by the same orientational correlation function, $C_k$.
This function is normally expected to to decay exponentially with
a characteristic length scale, $l_{p,x}$, 
\begin{equation}
C_k\equiv \langle \cos \theta_{k} \rangle= \mbox{e}^{-k a/l_{p,x}}.
\label{elpx}
\end{equation}

Planning ahead, we might be tempted to fit the
correlation function to an exponential function, e.g. by
linear regression of $\ln C_k$ {\em vs.} $k$, and simply calculate $l_{p,x}$
from the slope in this case. This is not such a good idea, however, because
we are disregarding one of the fitting parameters. It is a little optimistic
to think that the fitting would reproduce the exact value, $C_0=1$, as the
prefactor of the exponential (or the intercept $\ln C_0 = 0$ in the linear
regression). In order to take both parameters into account, it is better to
integrate
\begin{equation}
l_{p,\infty}= a \int_0^\infty C_0 \mbox{e}^{-k a/l_{p,x}} dk = C_0 l_{p,x} .
\label{elpinf}
\end{equation}
We see that when $C_0=1$ we get $l_{p,x}$ back and at the same time we have
a natural way to incorporate a fitted $C_0\neq 1$. The integral is essentially
the sum in Eq.~(\ref{elpi}) and here we have the connection between the
microscopic and exponential definitions. There is a difference, however.
The sum in Eq.~(\ref{elpi}) is truncated at the end of the chain regardless
if $C_k$ has decayed to zero or not, while the integral in Eq.~(\ref{elpinf})
extrapolates the result to that of an infinite chain. The significance of this
will become evident when we look at the results.

The correlation function $C_k$ is discrete since we are considering a polymer
with discrete bonds. We will get a continuous function, however, if we make
the bonds or segment lengths infinitesimally small, $a \rightarrow ds $,
and the number of segments infinitely large, while keeping the contour length,
$L=(N-1)a=\sum_{i=1}^{N-1}a=\int_0^L ds$, fixed. This is the worm-like chain model.\cite{Kratky:49,Yamakawa}
In the chain with discrete bonds there is automatically an intrinsic stiffness,
even in the
absence of interactions. This results, in the
well-known mean-square end-to-end distance for a freely jointed chain or
random walk with a fixed step length $a$,
\begin{equation}
R_{ee}^2=\sum_{i=1}^{N-1} a^2 C_0 + 2 \sum_{k=1}^{N-1} (N-1-k) a^2 C_k =
(N-1) a^2 ,
\label{eRflc}
\end{equation}
since there are $N-1-k$ scalar products of the type $\langle {\bf r}_i \cdot
{\bf r}_{i+k} \rangle$ [cf. Eq.~(\ref{eRee})] and without interactions one has
\begin{equation}
C_k = \left\{ \begin{array}{ll}
         1 & \mbox{for $k=0$} \\
         0 & \mbox{otherwise} \end{array}
            \right. .
\end{equation}
For the worm-like chain we need to specify $C(s)\equiv \langle \cos \theta(s)
\rangle$, where $s=\int_0^s ds$ is a distance along the contour of the chain, in order
to calculate [cf. Eq.~(\ref{eRflc})]
\begin{equation}
R_{ee}^2= 2\int_0^L \left( L - s \right) C(s) \  ds .
\end{equation}
Returning to the assumption of an exponential decay, i.e. $C(s)=\exp\left( 
-s/l_{p,w} \right)$, we get
\begin{equation}
R_{ee}^2= 2 L l_{p,w} - 2 l_{p,w}^2 \left( 1 - \mbox{e}^{-L/l_{p,w}} \right) .
\label{elpw}
\end{equation}

If $R_{ee}$ is known, $l_{p,w}$ can be obtained from \ref{elpw}. This method
is often used to extract an apparent persistence length from experimental
data,\cite{Schmidt:84,Reed:91a,Reed:91b} although what is usually measured is
the radius of gyration, $R_G$, defined as
\begin{equation}
R_G^2\equiv \frac{1}{N} \sum_{i=1}^N \langle \left( {\bf x}_i - {\bf x}_{cm}
\right)^2 \rangle = \frac{1}{N^2} \sum_{i=1}^N \sum_{j=i+1}^N \langle x_{ij}^2 \rangle,
\label{eRG}
\end{equation}
where ${\bf x}_{cm}$ is the centre of mass coordinate and $x_{ij}$ is the distance
between atoms $i$ and $j$. The worm-like chain expression used is
\begin{equation}
R_G^2= \frac{2}{L^2} \int_0^L ds^\prime \left( L-s^\prime \right)
\int_0^L ds \left(s^\prime - s \right) \mbox{e}^{-L/l_{p,w}} =
\frac{L l_{p,w}}{3} - l_{p,w}^2 + 2\frac{l_{p,w}^3}{L} -2\frac{l_{p,w}^4}{L^2}
\left( 1 - \mbox{e}^{-L/l_{p,w}} \right)
\label{elpwG}
\end{equation}
or its long chain limit ($L\gg l_{p,w}$)
\begin{equation}
R_G^2\approx \frac{L l_{p,w}}{3} .
\end{equation}
For a worm-like chain with an exponentially decaying correlation function,
Eqs. (\ref{elpw}) and (\ref{elpwG}) are equivalent. In other cases the obtained
persistence lengths would hardly be identical, but the trends are expected to be
the same. For convenience we will be using Eq. (\ref{elpw}). 

The persistence lengths discussed above are all
total persistence lengths, where the "bare" part, the persistence
length 
in the absence of electrostatic interactions, depends on the
chain (model) itself. For example, for the freely jointed chain $l_{p,0}$ is equal
to the bond length [cf. Eqs.~(\ref{elpR}) and (\ref{eRflc})] and,
experimentally,
a value extrapolated to infinite amounts
of salt is often used. In order to calculate the electrostatic part, the OSF
theory considers the electrostatic contribution
to the bending energy, which is quadratic in the curvature or bending
angle, for a very stiff worm-like chain ($l_{p,0} \gg
\kappa^{-1}$), whose electrostatic interactions are described by
the Debye-H\"{u}ckel approximation, i.e. a screened Coulomb potential
\begin{equation}
u_{sc}(r)=k_B T l_B\frac{\mbox{e}^{-\kappa r}}{r} ,
\label{eusc}
\end{equation}
where $k_B$ is Boltzmann's constant, $T$ is the absolute temperature and 
$l_B=e^2/(4\pi \epsilon_{r} \epsilon_{0} k_{B} T)$ is the Bjerrum length,
with $e$ the elementary charge, $\epsilon_{r}$ the dielectric constant
of the solution and $\epsilon_{0}$ the permittivity of vacuum. The
Debye screening length, $\kappa^{-1}$, is given by $\kappa^2=8 \pi l_B N_A I 1000$,
where $N_A$ is Avogadro's number and $I$ is the ionic strength in units
of molar concentration (SI units are implied for the other quantities). The
Debye-H\"{u}ckel approximation is expected to be reasonable for a 1:-1 salt
(and a weakly charged polyelectrolyte),
in which case the ionic strength is just the molar salt concentration, i.e.
$I=C_s$.
The predicted electrostatic persistence length is
\begin{equation}
l_{OSF}=\frac{l_B Z^2}{12} \left[ 3y^{-2}-8y^{-3}+\mbox{e}^{-y} (y^{-1}+
5y^{-2}+8y^{-3}) \right],
\label{elOSF}
\end{equation}

with $y\equiv \kappa L$ and $Z$ the amount of elementary charges on the chain.
For large $y$ this reduces to
\begin{equation}
l_{OSF}^\prime = \frac{l_B}{4 \kappa^2 A^2},
\label{elOSFp0}
\end{equation}
where $A=L/Z$ is the length per unit charge. If we translate this to a model
with a certain bond length $a$ and a charge per monomer $\alpha=Z/N$ we get
\begin{equation}
l_{OSF}^\prime = \frac{\alpha^2 l_B}{4 \kappa^2 a^2}.
\label{elOSFp}
\end{equation}

Besides the bond interactions, the models used in the simulations of
the flexible polyelectrolyte chains only
contain electrostatic interactions, which allows us to obtain a truly
{\em electrostatic} persistence length. 
In the simulations of the stiff polyelectrolyte chains we also include
a bond angle term. In a real polyelectrolyte other
interactions, like excluded volume and attractive interactions, may be
significant. Still, what we see here is in very good agreement with experiment.

\section{MODELS}
The polyelectrolyte is regarded as an infinitely diluted polyacid in a
salt solution. The monomers form a linear chain where each monomer
represents a charged site that carry one unit of negative charge.
Three different models have been considered: Model 1 and 2 are freely jointed
chains with harmonic and rigid bonds, respectively. In model 3 the rigid bonds
have been augmented with an angular potential restricting the orientation of
adjacent bonds.
The harmonic bonds correspond to an approximately Gaussian distribution of bond lengths
and can be viewed as an averaging over the conformational freedom of the
covalent structure that implicitly joins the charged sites.

The solvent is treated as a dielectric continuum with a dielectric constant
$\epsilon_{r}=78.3$, corresponding to water at room temperature, and the
salt ions are implicitly represented by the screening parameter $\kappa$.
Thus, charged monomers interact via a screened Coulomb potential, $E_{C}$,
which for the flexible
bond model (model 1) is augmented with a Gaussian term, $E_{G}$, leading to a total
interaction energy
\begin{equation}
  E=E_{G}+E_{C}=\frac{k}{2} \sum_{i \not=N } r_{i,i+1}^{2} +
 \frac{\alpha^2 e^{2}}{4 \pi \epsilon_{r} \epsilon_{0}} \sum_{i}
 \sum_{j>i} \frac{\mbox{e}^{-\kappa r_{i,j}}}{r_{i,j}} ,
\label{etoten}
\end{equation}
where $N$ is the number of monomers, $r_{i,j}$ is the distance between
monomer $i$ and $j$, $e$ is the
electronic charge, 
$\epsilon_{0}$ is the permittivity of
vacuum and $\alpha$ the (fractional) amount of charge on a monomer $i$.
The force constant,
$k$, is given implicitly through the input parameter $r_{0} =
 (e^{2}/4 \pi \epsilon_{r} \epsilon_{0} k)^{1/3}$, which is the equilibrium
distance for a fully charged dimer.

The angular potential in model 3 is simply,
\begin{equation}
  E_{A}=-k_{B}T\frac{l_{B}l_{A}^{2}}{r_{0}^{3}} \sum_{i=1}^{N-1}
 \cos \theta_{i} ,
\label{eang} 
\end{equation}
where $\theta_{i}$ is the angle between bond $i$ and $i+1$. The angular potential is 
characterized by the length $l_A$ and increasing 
$l_A$ means increased stiffness in the chain.

\section{METHODS}
The key approach employed for studying the scaling properties of the persistence
length is the Monte Carlo method. In this section, however, we also discuss what is
expected from non-stochastic approaches; a variational method, high- and low-T
expansions.

\subsection{Monte Carlo simulation}
The Monte Carlo (MC) simulations were performed with the traditional Metropolis
al\-go\-rithm\cite{Metropolis:53} in a canonical ensemble. One
polyelectrolyte chain at infinite dilution was simulated.
The sampling of chain conformations was made highly efficient, compared to
traditional Monte Carlo moves, by using a pivot algorithm, which allows
chain lengths of more than 2000 monomers. The pivot algorithm was first
described by Lal,\cite{Lal:69} and its efficiency for self-avoiding walks
have been thoroughly discussed by Madras and Sokal.\cite{Madras:88}
Despite its excellent properties, it was only recently that the pivot algorithm
started to gain wider recognition as a tool to enhance simulations of polymers
and polyelectrolytes.\cite{Bishop:91,Seidel:94,Jonsson:95a,Ullner:94}

In a traditional single-move algorithm only one or a few monomers are translated
or rotated at a time.
The number of interactions that
has to be calculated is of the order $N$ and a
large number of attempts per monomer is needed to generate independent chain
conformations. Similarly, in the pivot algorithm each monomer $i$ (except
the first one) is translated in turn but together with the rest of the chain
(monomers $i+1$ to $N$). Furthermore, the last part of the polymer is then
rotated as a rigid body around one of the coordinate axes with monomer $i$ as
origin. To be specific, if the bonds are flexible (model 1) they are both
translated and rotated, while in the case of rigid bonds (models 2 and 3) only
the rotation
is performed. The number of interactions calculated in one step is of the order
$N^{2}$ but independent conformations are obtained after a only a few attempted
moves, on the order of one per monomer or $N$ in total. The net effect is a
greatly reduced simulation time for a given degree of precision and a
computational cost that grows approximately as $N^{3}$.

One expects the pivot algorithm to have its maximal efficiency
for unscreened and extended chains, but we have recently shown its efficiency
even for screened chains with less elongated structure.\cite{Ullner:96a} The pivot algorithm
provides an effective sampling of global properties like $R_{ee}$, but it turns
out to be superior also for local properties like the monomer-monomer separation.
A more detailed investigation than the present, is needed to
really map out the applicable range for the pivot algorithm.

\subsection{Variational approach}
In the variational approach the true Boltzmann distribution $P
\propto\exp(-E/k_BT)$ is approximated by a trial distribution $P_V
\propto\exp(-E_V/k_BT)$, having a set of free parameters.  It yields a
variational free energy
\beq    \hat{F} = \la E \ra_V - k_BT S_V \geq F,
\eeq
bounded from below by the true free energy $F$.  $\la E \ra_V$ is the
average of the true energy in the trial Boltzmann distribution $P_V$
and $S_V$ is the corresponding variational entropy.  By minimizing
$\hat{F}$ with respect to the parameters of $E_V$, thermodynamic
quantities can be estimated.  In the case of model 1 (flexible
chain), we consider an arbitrary pure Gaussian variational
distribution,\cite{Jonsson:95a} corresponding to
\beq    E_V/k_BT = \half \sum_{ij} \Ginvij \ri\cdot\rj ,
\eeq
where $\ri$ denotes the $i$th bond vector.  With the correlation matrix $G$
optimized (numerically) so as to minimize $\hat{F}$, the variational
estimate for the correlation matrix becomes
\beq    \la \ri\cdot\rj \ra_V = 3\Gij.
\eeq
From this a variational estimate for the persistence length, e.g. as
defined in Eq. (\ref{elpR}), can be obtained.

\subsection{High temperature expansion}
For model 1 at high temperatures, the energy [Eq. (\ref{etoten})]
will be dominated by the bond term, and the interaction $E_C$ can be
treated as a perturbation.  A perturbative expansion for the thermal
average of an arbitrary observable $f$ yields
\beq    \la f \ra = \la f \ra^0 - \frac{1}{k_BT}\la f E_C\ra^0_C
        + \frac{1}{2(k_BT)^2} \la f E_C E_C\ra^0_C + \ldots ,
\label{ehT}
\eeq
where $\la\,\ra^0_C$ refers to connected expectation values in the
unperturbed Boltzmann distribution.

In a continuum approximation, valid when the nearest-neighbour
distance is small as compared to the range of the interaction, the
first few terms in the perturbative expansion of the squared
end-to-end distance $R_{ee}^2$ are given by \cite{Soederberg:96} (assuming full
charge, $\alpha=1$)
%
\beq
        \la R_{ee}^2\ra =
        \frac{N r_0^3}{l_B}
        \left [ 3
        + \gamma f_1(\mu)
        + \gamma^2 f_2(\mu)
        + \ldots \right ] .
\eeq
where
\beq
        f_1(\mu) = 4 \sqtopi - 16 \mu^{-1/2} + {\cal O}(\mu^{-1})
        \; , \; \;
        f_2(\mu) = -\(\frac{32}{\pi} - \frac{56}{9}\) + {\cal O}(\mu^{-1/2})
\eeq
with the dimensionless parameters $\mu$ and $\gamma$ given by
\beq
        \mu = \frac{N \kappa^2 r_0^3}{l_B}
        \; , \; \;
        \gamma = \frac{ N^{1/2} l_B^{5/2} }{ \kappa^2 r_0^{9/2} } .
\eeq
Note that $\mu$ is proportional to the squared unperturbed end-to-end
distance in units of the screening length; in the parameter range we
are studying, $\mu$ is large.

Keeping only the leading large-$\mu$ contribution to each term $f_i$
yields a power series in $\gamma$, indicating that the expansion is
valid for small $\gamma$. Unfortunately, for the parameter range we
are studying $\gamma$ is {\em not} small, and while the first order
expansion is proportional to $\kappa^{-2}$, the next order term, proportional
to $\kappa^{-4}$, is larger and has a negative value.

Similarly, for the average squared bond length we obtain in the continuum
approximation the perturbative result
\beq
        a^2 \equiv \la r_{mm}^2\ra =
        \frac{r_0^3}{l_B}
        \left [ 3
        + 4 \frac{l_B^2}{\kappa r_0^3}
        + \ldots
        \right ] .
\eeq

The persistence length $l_{p,R}$ is then obtained by dividing
$R_{ee}^2$ by $N r_{mm}$. In the results section, we have included the
first order perturbative result for comparison, although a good
agreement cannot be expected since the parameter $\gamma$ is not
small.

\subsection{Low temperature expansion}
A similar approach can be followed for low temperatures, where the
Boltzmann distribution is dominated by the minimal energy
configuration. For small fluctuations around the ground state, the
energy can to lowest order be considered Gaussian, with the
corrections treated as a perturbation. The resulting correlation
matrix $\la\ri\cdot\rj\ra$ \cite{Peterson:96} can be used to calculate
the persistence length.

\section{RESULTS AND DISCUSSION}
\subsection{Flexible polyelectrolyte}
\begin{figure}[t]
\hspace*{2.5cm} \hbox{
        \psfig{figure=\PLOTDIR/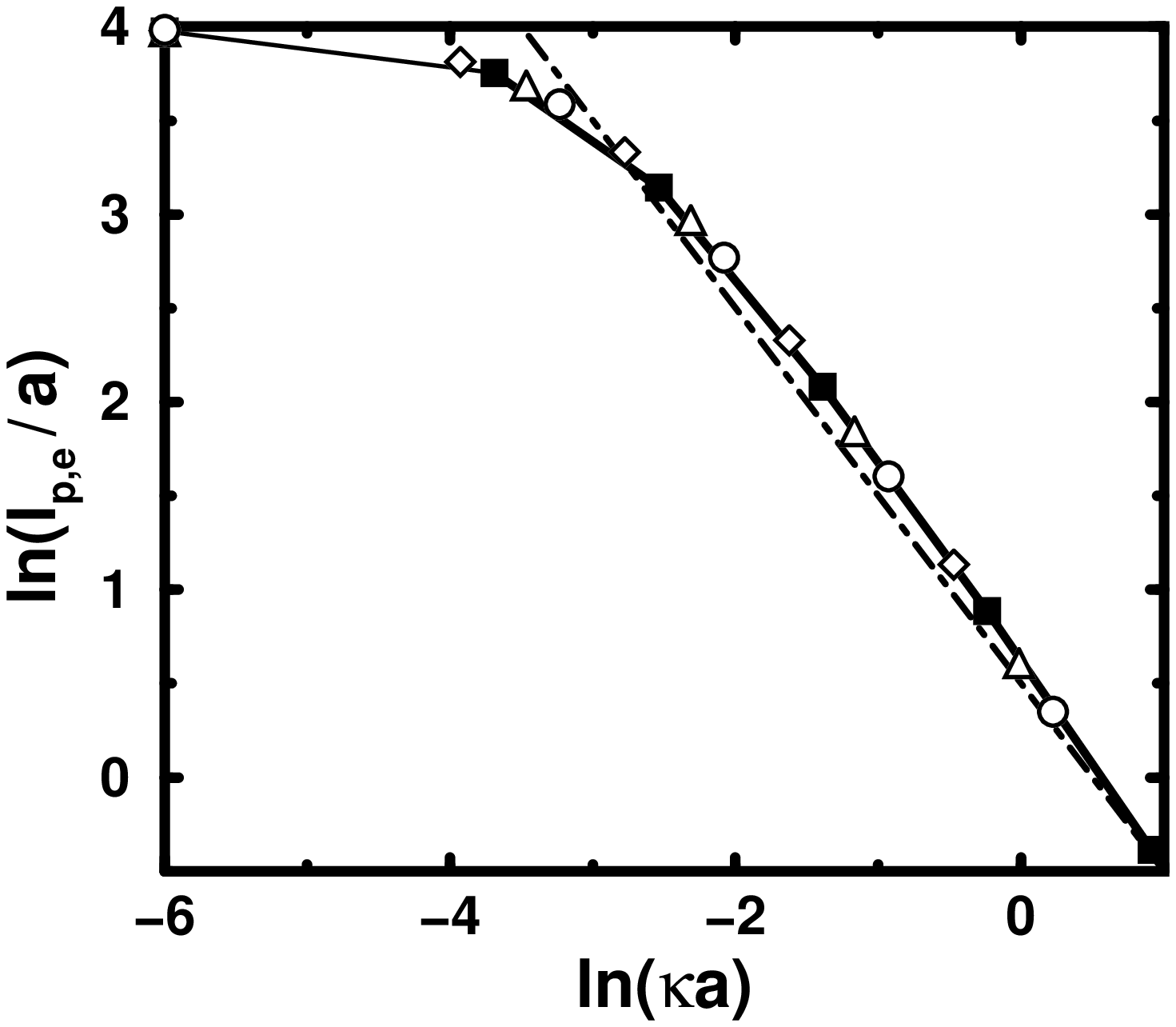,width=9.15cm}
}
\vspace{-2.5cm}
\caption{The logarithm of the scaled electrostatic persistence length,
         where $l_{p,e}= l_{p,R}-a$ [cf. Eq.~(\protect\ref{elpR})],
         as a function of $\ln(\kappa a)$ for rigid bond chains with $N=320$ and
         fixed $\xi_p$. Filled squares represent $a=24$ \AA\ and $\alpha=1$ and the
         open symbols represent $\alpha$/$a=0.3536$/3 \AA\ (triangles), 0.5/6 \AA\
         (diamonds) and 0.707/12 \AA\ (circles),  with the values on the y-axis
         showing the results for the salt-free case.
         The dot-dashed line marks a slope of -1.
\label{fscb}
}
\vspace{1.5cm}
\hspace*{2.5cm} \hbox{
        \psfig{figure=\PLOTDIR/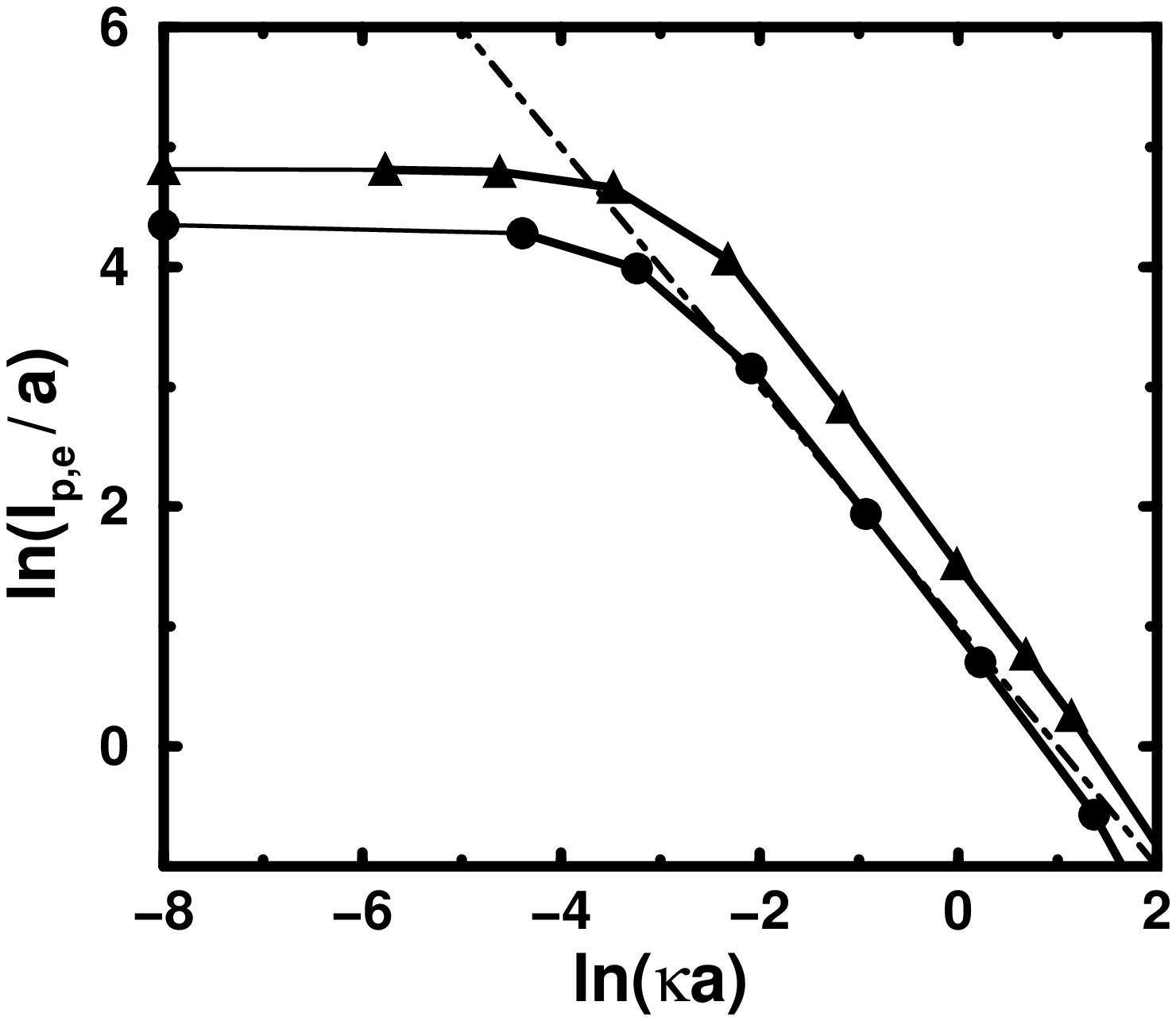,width=9.15cm}
}
\vspace*{-2.5cm}
\caption{The logarithm of the scaled electrostatic persistence length,
         where $l_{p,e}= l_{p,R}-a$
         [cf. Eq.~(\protect\ref{elpR})], as a function of $\ln(\kappa a)$.
         The results are for rigid bond chains with $N=320$
         and $a=3$ \AA\ (triangles) and 12 \AA\ (circles) and fully charged monomers.
         The values on the y-axis are for the salt-free case, and are
         connected to the rest of the points by thinner lines. The dot-dashed
         line denotes a slope of -1.
\label{fsca}
}
\end{figure}

From the form of the Hamiltonian, being just a sum of screened Coulomb interactions
[Eq.~(\ref{eusc})], and the fact that the statistical mechanical weight of a
conformation is given by $\exp(-E/k_{B}T)$,
it is easy to see that the rigid bond
model scales with the bond length $a$ as the unit of length and that any
scaled length, e.g. $l_p/a$, must be a function of the dimensionless numbers
(scaled length scales) $\kappa a$ and the strength parameter
\begin{equation}
\xi_p=\frac{\alpha^2 l_B}{a},
\label{exip}
\end{equation}
which measures the strength of the electrostatic interactions within
the polyelectrolyte for a given $N$. As a visual proof,
Fig.~\ref{fscb} shows that $l_{p,e}/a$ is constant for a given value of these
parameters.
Note that $\xi_p$ is different from the
so called Manning parameter,\cite{Manning:69} $\xi=l_B/A=\alpha l_B/a$, which
measures the interactions between the polyelectrolyte and mobile ions.
In a previous paper we found for a titrating polyelectrolyte that the
approximation of putting full charges a distance $A$ apart is inferior to
having fractional charges on the actual titrating sites.\cite{Ullner:96b}
Thus, we may conclude that $\xi_p$ is the more pertinent parameter in this
context.

The results in Fig. \ref{fsca}
show two distinct regimes in the
salt dependence of the electrostatic persistence length.
For low salt concentrations, i.e. when the Debye screening length is longer
than the dimensions of the chain ($ \kappa^{-1} \gg R_{ee}$), $l_{p,e}$ is almost
constant. Note that the symbols on the y-axis represent the values in the
unscreened case ($\kappa=0$). When $\kappa^{-1} \ll R_{ee}$ the relationship
between $l_{p,e}$ and $\kappa^{-1}$ is close to linear. There is actually a third regime when $\kappa^{-1}$ is less than the
bond length, but we will save that till the  discussion of the variational
results.

\begin{figure}[t,b,p]
\hspace{2.5cm} \hbox{
        \psfig{figure=\PLOTDIR/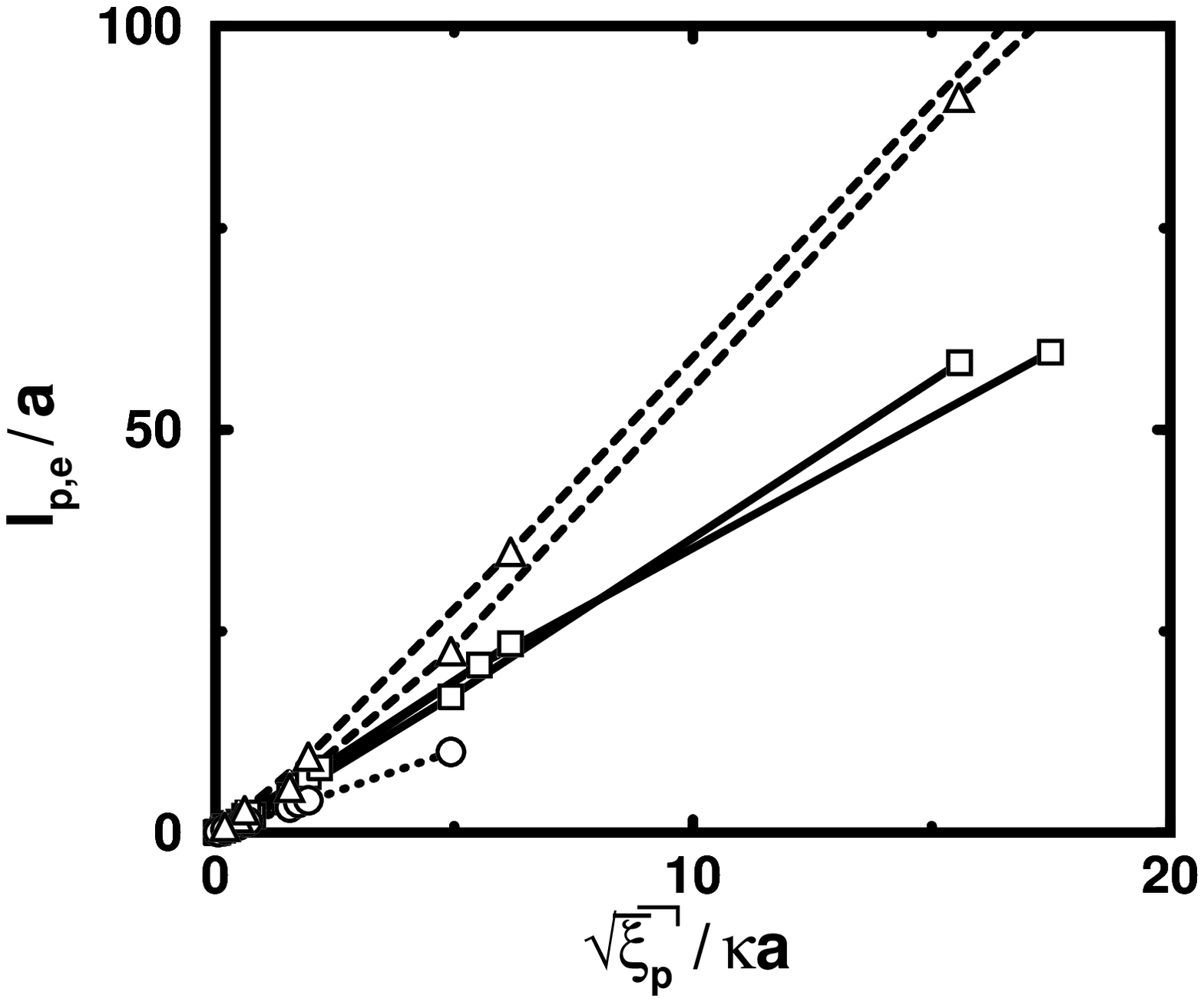,width=9.15cm}
}
\vspace{-2.5cm}
\caption{The scaled electrostatic persistence length, where
         $l_{p,e}=l_{p,R}-a$, as a function of $\protect\sqrt{\xi_p} / \kappa a$.
         The results are for rigid bond chains with $N=80$ (circles) and 320
         (squares) with bond lengths 3 \AA, 6 \AA, 12 \AA\ and 24 \AA\
         as well as $N=1000$ with $a=3$ \AA\ and 12 \AA\ (triangles). In order to clearly
         show the linear regime, only points where
         $\kappa^{-1}\leq R_{ee}/10$ are displayed.
\label{fN0}
}
\vspace{1.5cm}
\hspace{2.5cm} \hbox{
        \psfig{figure=\PLOTDIR/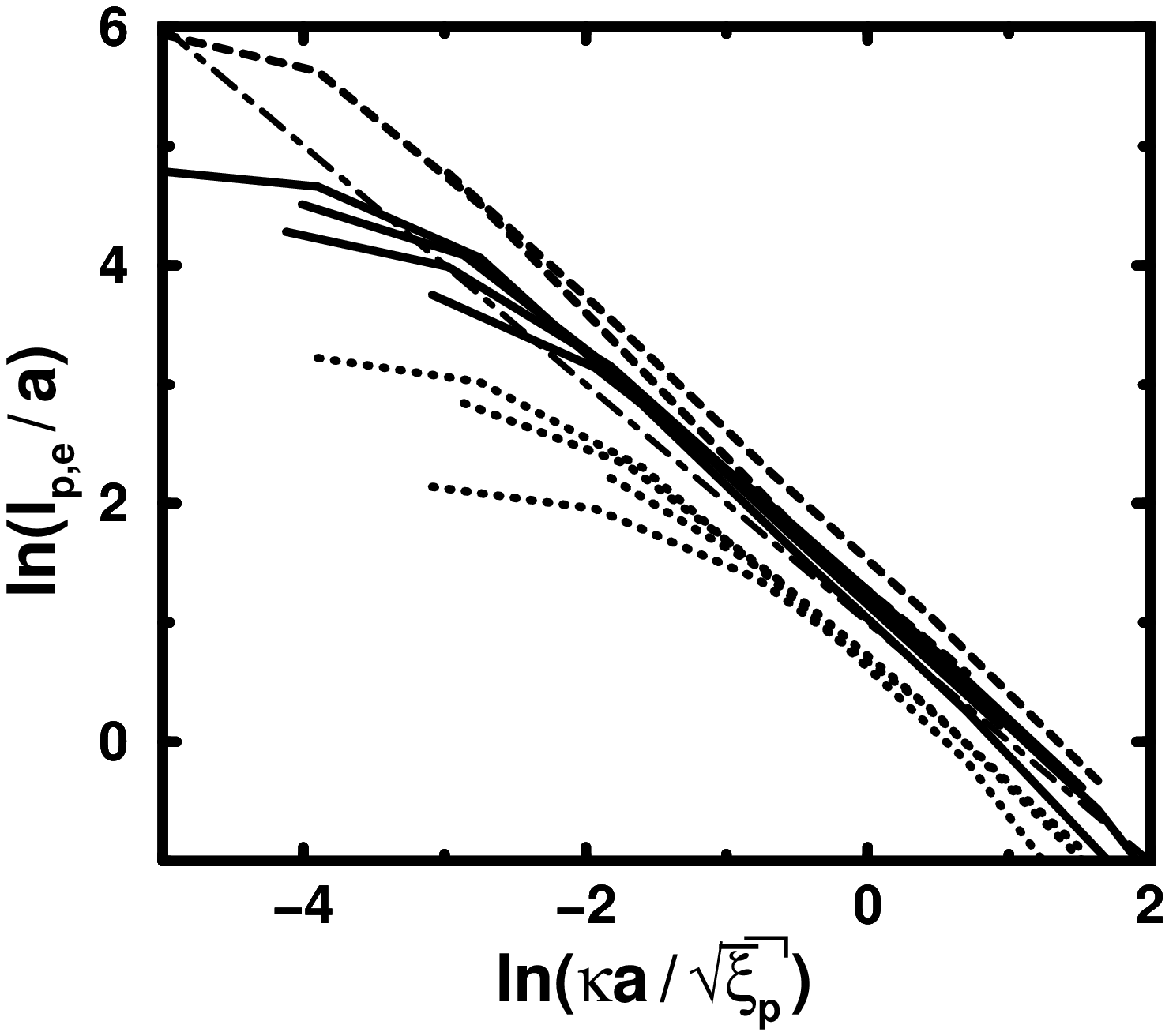,width=9.15cm}
}
\vspace{-2.5cm}
\caption{The logarithm of the scaled electrostatic persistence length,
         where $l_{p,e}=l_{p,R}-a$, as a function of $\protect\sqrt{\xi_p}/\kappa a$.
         The lines represent the same combinations of $a$ and $\alpha$ as
         in \protect\ref{fN0}.
         The dot-dashed line marks a slope of -1.
\label{fN}
}
\end{figure}

Experimentally, Reed {\em et al.} have found a linear dependence on both
$\kappa^{-1}$ and $\alpha$ for $\l_{p,e}$.\cite{Reed:91b} To us, this implies
\begin{equation}
\l_{p,e}\sim \frac{\sqrt{\xi_p}}{\kappa}=\frac{\alpha}{\kappa}
\sqrt{\frac{l_B}{a}}=\frac{\alpha}{\sqrt{8000\pi N_A I a}}
\label{elpsc}
\end{equation}
at high salt concentrations,
which is confirmed in Figs. \ref{fN0} and \ref{fN}.
Note that $l_{p,e}$ is predicted to be temperature independent.
Ha and Thirumalai (HT) have derived a similar expression,
$l_p \sim \sqrt{(l_{p,0} l_B/A^2)}/\kappa$
for $l_{OSF} \gg l_{p,0}$.\cite{Ha:95} As long as $l_{p,0}=a$ the two expressions are
the same, although $l_p$ in their case appears to be the total persistence
length. However, if we add a repulsive potential between next-nearest 
neighbours, which increases $l_{p,0}$ (but not too much, see below),
Eq.~(\ref{elpsc}) still holds, while the
HT expression with the actual intrinsic persistence length does not.
The slope for the highly screened regime ($ \kappa^{-1} \ll R_{ee}$) in Fig. \ref{fN}
is roughly -1.1.

Equation (\ref{elpsc}) has no $N$ dependence. Naively one may expect
excluded volume behaviour in the high salt regime as a result of short ranged
electrostatic interactions between monomers, which can also be far apart in the
sequence
as the chain loops back on itself. For a self-avoiding chain we have
$R_{ee}^2\sim N^{1.2} a^2$,\cite{Flory} which from Eq.~(\ref{elpR}) would
suggest $l_{p,e}\sim N^{0.2}$ for long chains. The slopes of the linear
parts of Fig. \ref{fN0}, roughly 2.6, 4.3 and 7.3 for $N=80$, 320 and 1000,
respectively, give an $N$-exponent of 0.4.

\begin{figure}[t,b,p]
\hspace{2.5cm} \hbox{
        \psfig{figure=\PLOTDIR/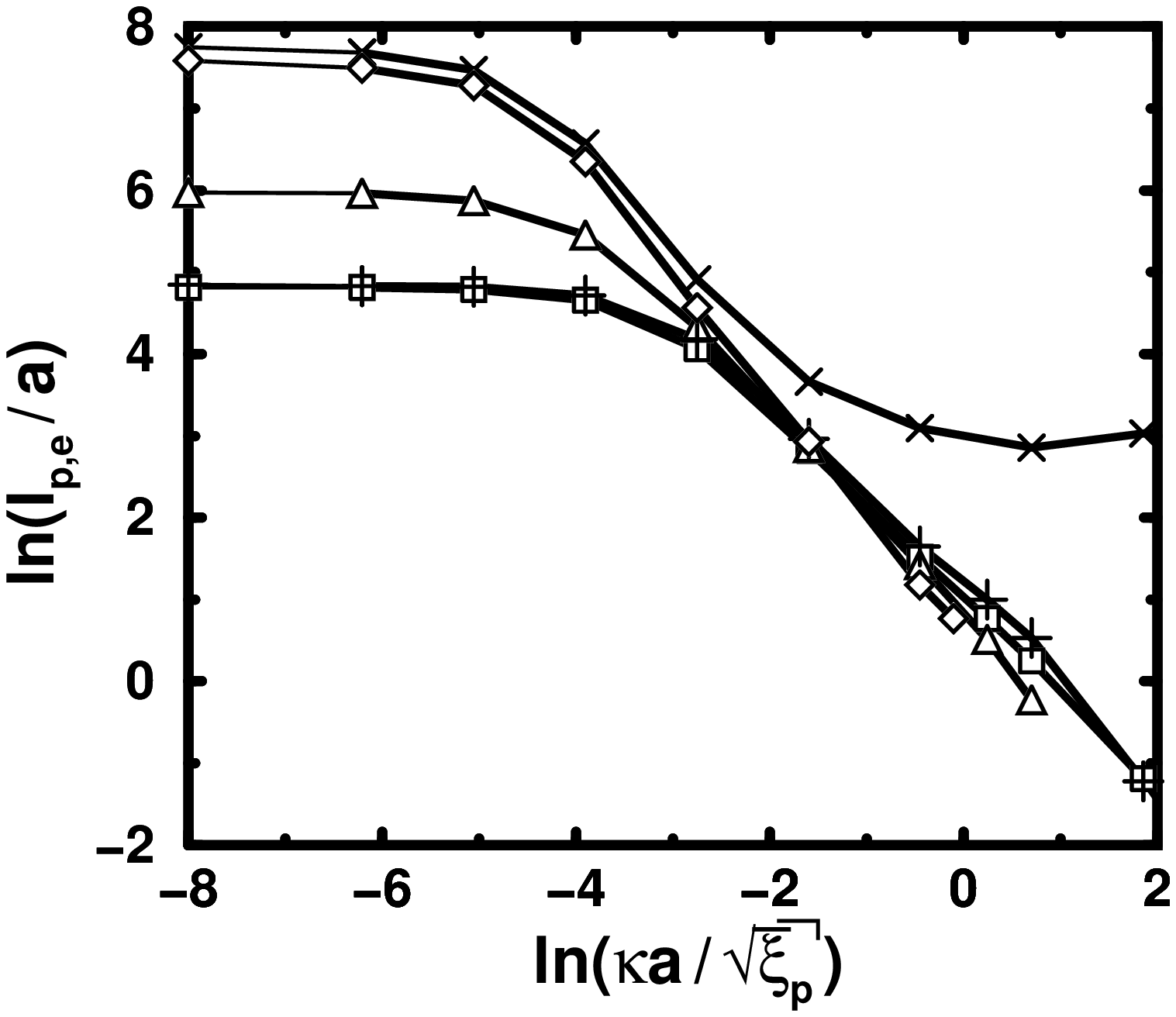,width=9.15cm}
}
\vspace{-2.5cm}
\caption{The logarithm of the electrostatic persistence length
         as a function of the logarithm of $\kappa a/ \protect\sqrt{\xi_p}$
         for different
         definitions of $l_{p,e}$: $l_{p,N/2}-a$ from Eq. (\protect\ref{elpi}) (plus signs),
         $l_{p,R}-a$ from Eq. (\protect\ref{elpR}) (squares),
         $l_{p,x}-a$ from Eq. (\protect\ref{elpx}) (crosses),
         $l_{p,\infty}-a$ from Eq. (\protect\ref{elpinf}) (diamonds),
         $l_{p,w}-a$ from Eq. (\protect\ref{elpw}) (triangles).
         The values are based on simulation results, either directly or 
         through $R_{ee}$, of 320-mers with $a=3$ \AA. 
}
\label{fdef}
\end{figure}

So far we have used the ``macroscopic'' persistence length,
Eq. (\ref{elpR}), but the same conclusions are reached with 
other definitions as can be seen in Table \ref{tall} and Fig. \ref{fdef}.
\begin{table}[t,b,p]
\caption{End-to-end separation and electrostatic persistence lengths from
different definitions, $l_{p,N/2}-a$ from Eq. (\protect\ref{elpi}),
         $l_{p,R}-a$ from Eq. (\protect\ref{elpR}),
         $l_{p,x}-a$ from Eq. (\protect\ref{elpx}),
         $l_{p,\infty}-a$ from Eq. (\protect\ref{elpinf}),
         $l_{p,w}-a$ from Eq. (\protect\ref{elpw})  and
         $l_{OSF}$ from Eq. (\protect\ref{elOSF}), obtained for a 320-mer.
         The unit of length is 1 \AA.}
\begin{center}
\begin{tabular}{ccrrrrrrrr}
\hline \hline
\multicolumn{1}{c}{$a$} & \multicolumn{1}{c}{$C_s$/M} &
\multicolumn{1}{c}{$\kappa^{-1}$} &  \multicolumn{1}{c}{$R_{ee}$} &
\multicolumn{1}{c}{$l_{p,N/2}-a$} & \multicolumn{1}{c}{$l_{p,R}-a$} &  
\multicolumn{1}{c}{$l_{p,x}-a$} & \multicolumn{1}{c}{$l_{p,\infty}-a$} &
\multicolumn{1}{c}{$l_{p,w}-a$} & \multicolumn{1}{c}{$l_{OSF}$}   \\ \hline
   & 0 &        ---   & 844 & 383 & 371 & 6969 & 5909 & 1182 & 10185 \\
   & 0.00001 & 961 & 843 & 376 & 370 & 6494 & 5412 & 1169 &  9191 \\
   & 0.0001 &  304 & 834 & 372 & 362 & 5246 & 4365 & 1071 &  5655\\
   & 0.001 &    96 & 781 & 334 & 317 & 2152 & 1713 &  699 &  1353 \\
 3 & 0.01 &     30 & 581 & 197 & 175 &  409 &  288 &  228 &   169 \\
   & 0.1 &       9.6 & 316 &  58 &  51 &  117 &  56  &   52 &    18 \\
   & 1 &         3.0 & 171 &  16 &  14 &   66 &  10  &   12 &     2 \\
   & 10 &        0.96 & 101 &   5 &   4 &   52 &  -1 &     2 &     0.2 \\
   & 100      &   0.30 & 68 & 0.9 & 0.9 &   62 &  -3 &  -0.6 &     0.02 \\
\hline
   & 0 &        ---   & 2680 &1018 & 932 & 9158 & 5636 & 1433 & 10185 \\
   & 0.00001 & 961 & 2586 & 943 & 867 & 7258 & 4231 & 1267 & 4605 \\
   & 0.0001 &  304 & 2236 & 713 & 647 & 3300 & 1678 &  821 &  910 \\
   & 0.001 &    96 & 1483 & 324 & 281 &  978 &  378 &  301 &  108 \\
12 & 0.01 &     30 &  828 &  99 &  84 &  461 &   87 &   80 &   11 \\
   & 0.1 &     9.6 &  481 &  27 &  24 &  256 &   14 &   18 &    1 \\
   & 1 &       3.0 &  313 &   7 &   7 &  213 &  -11 &  0.8 &  0.1 \\
   & 10 &     0.96 &  234 &   4 &   1 &  262 &  -12 &   -5 & 0.01 \\
   & 100 &    0.30 &  217 &   1 & 0.2 &  420 &  -12 &   -6 &0.001 \\
\hline \hline
\end{tabular}
\end{center}
\label{tall}
\end{table}
The two persistence lengths obtained directly form the simulation,
$l_{p,N/2}$ and $l_{p,R}$ [Eqs.~(\ref{elpi})
and (\ref{elpR})], are almost identical. $l_{p,R}$ is a global quantity
and thus have better statistics than $l_{p,N/2}$, which has more local
character. This becomes noticeable at high salt concentrations where the
latter starts to suffer from insufficient
averaging.
The worm-like chain
persistence length, Eq.~(\ref{elpw}), also give the same results
for $\kappa^{-1} \ll R_{ee}$ and $\kappa^{-1}  > a$. For screening lengths less than the bond length the true
structure of the chain becomes important and the worm-like chain model breaks
down and even gives negative persistence lengths. To save the worm-like chain,
a term corresponding to the last term of Eq.~(\ref{elpR}) would have been
needed in Eq.~(\ref{elpw}).

$l_{p,x}$ [Eq.~(\ref{elpx})] and $l_{p,\infty}$ have been obtained by fitting
an exponential to the orientational correlation function. In order to avoid
end-effects we have calculated $C_k$ from the central bond and fitted the
function only in the range $k$=0--120 ($N$=320).
\begin{figure}[t,b,p]
\vspace*{1cm}
\hspace*{2.5cm} \hbox{
        \psfig{figure=\PLOTDIR/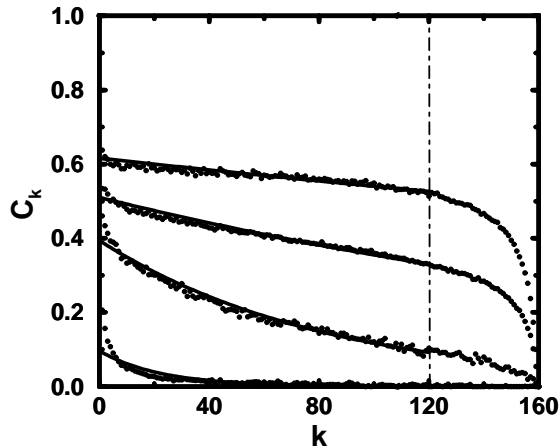,width=9.15cm}
}
\vspace*{-2.5cm}
\caption{The orientational correlation function for a 320-mer with $a=12$ \AA\
         and salt concentrations of 0, 0.1, 1, 100 mM (top to bottom)
         represented by small circles.
         The solid lines are exponentials fitted
         in the range $k$=0--120. The dot-dashed line just shows the upper
         limit of the fitting and the upper limit of the y-axis is a reminder
         that $C_0=1$.
}
\label{flpx}
\end{figure}
Figure \ref{flpx} shows the correlation
function together with the fitted exponential. It is clear that some
information is lost at small values of $k$ and that this is of greater
importance at higher salt concentrations.
In particular the initial part at $C_0=1$, which for the
integrated persistence length $l_{p,\infty}$ corresponds to the
intrinsic persistence length, $l_{p,0}=a$, is lost. Table \ref{tall}
shows that if we did not subtract $l_{p,0}$ from $l_{p,\infty}$, this
persistence length would be in good agreement with the microscopic
definitions, not only at intermediate salt concentrations, but also at
higher salt concentrations.
Even more information is lost when calculating $l_{p,x}$.
Note for example that the fitted
exponentials have a prefactor ($C_0$) that is less than one and decreasing faster than
$C_1$. In Fig.~\ref{fdef} we see that $l_{p,x}$ has a behaviour of its own.
There is not a simple, consistent $\kappa^{-1}$ dependence and an apparent
exponent $p$ in $l_{p,x}\sim\kappa^{-p}$ can have a range of values less
than one, which is exactly the result of Micka and Kremer.\cite{Micka:96}
Their conclusions are thus a consequence of an unfortunate choice of persistence
length definition.

\begin{figure}[t,b,p]
\hspace{2.5cm} \hbox{
        \psfig{figure=\PLOTDIR/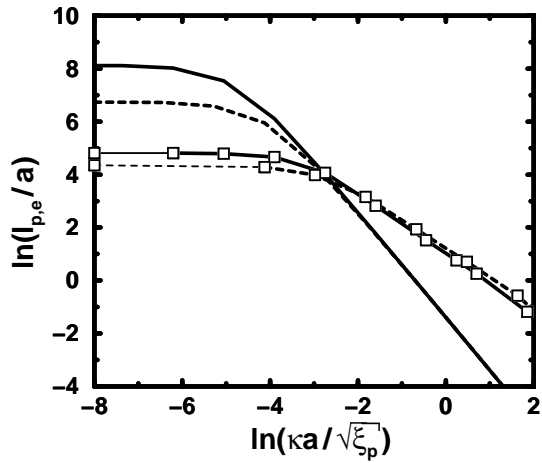,width=9.15cm}
}
\vspace{-2.5cm}
\caption{OSF theory [Eq. (\protect\ref{elOSF})] (no symbols) compared
         to the simulated electrostatic persistence length, $l_{p,e}=l_{p,R}-a$
         from Eq. (\protect\ref{elpR}), (squares).
         The results are for 320-mers with $a=3$ \AA\ (solid lines) 
         and $a=12$ \AA\ (dashed lines).
         The values on the y-axis
         are the results for the salt-free case.
}
\label{fOSF}
\end{figure}
The OSF persistence length also has an almost constant
regime, with a salt-free limit $l_{OSF}=l_B Z^2/72$ (see Fig.~\ref{fOSF}).
In this region
the value of the electrostatic persistence length is sensitive to whether it
includes end-effects, like $l_{p,i}$ and $l_{p,R}$, or is defined through
a correlation function that is extrapolated to infinite chain lengths,
like $l_{p,\infty}$ (and in principle $l_{p,x}$ if $C_0$ is close to one,
cf. Eq.~(\ref{elpinf})). $l_{OSF}$ belongs to
the latter group and is actually in fair agreement with 
$l_{p,\infty}$ (and $l_{p,x}$) in this regime (see Table \ref{tall}).
The worm-like chain expression, Eq.~(\ref{elpw}), gives a behaviour intermediate
between the two extremes and since
the worm-like chain is often used to
obtain persistence lengths from experimental data this is the behaviour
expected to be reported experimentally.
Such a report by F\"{o}rster {\em et al.} appears to be
at odds with both OSF theory and the notion of a power law
for the $\kappa$ dependence,\cite{Foerster:92} but a comparison with the
results here suggests that their experimental system just displays the constant
regime and
the subsequent cross-over region. This means that their comparison with
OSF theory is rather unfair since they are using the high-salt version,
Eq.~(\ref{elOSFp}). 

In the second regime, $l_{OSF}$ of course displays the well-known quadratic
dependence on the Debye length, in disagreement with simulations. It is
interesting to note that $l_{OSF}^\prime/a$ has the correct relationship
between $\xi_p$ and $\kappa a$. It is essentially the square of the
dimensionless form of the right-hand side of Eq.~(\ref{elpsc}). This is a 
little bit tricky, however. Going from Eq.~(\ref{elOSFp0}) to Eq.~(\ref{elOSFp})
the choice of the bond length $a$ is arbitrary, because it is always
balanced by $\alpha$. This is no longer true if we are to scale $l_{OSF}^\prime$
by $a$. In that case we need the proper rigid bond model to get the correct
bond length dependence, which is closely related to the above discussion about
the difference between having an effective charge or an effective bond
length, {\em i.e.} using $\xi_p$ or the Manning parameter, $\xi$.

\begin{figure}[t,b,p]
\hspace{2.5cm} \hbox{
        \psfig{figure=\PLOTDIR/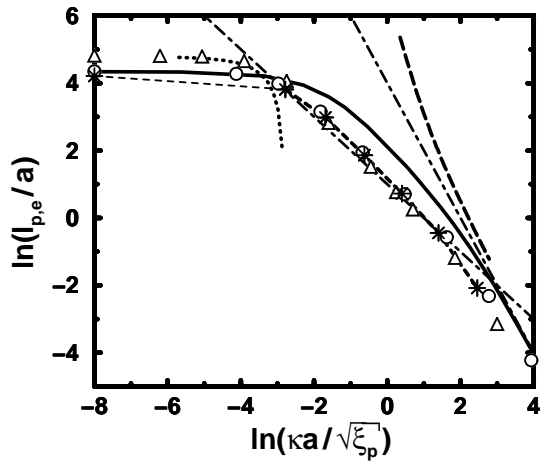,width=9.15cm}
}
\vspace{-2.5cm}
\caption{Comparison between simulations and variational and perturbational
         calculations.
         The simulation results are for 320-mers with rigid bonds
         of 3 \AA\ (triangles) and 12 \AA\ (circles) and flexible bonds
         with $r_0=6$ \AA\ (dashed line, stars). The solid line represent
         the variational result for the latter, while the dotted line
         and the long dashed lines are low temperature and high temperature
         expansions, respectively, for a 512-mer with $r_0=6$ \AA.
         The {\em symbols} on the y-axis
         are for the salt-free case and the dot-dashed lines mark a slope of
         -1 and -2, respectively.
}
\label{fvar}
\end{figure}

Figure \ref{fvar} shows that the electrostatic persistence length obtained
from the variational calculations is slightly higher than the simulation
results for the same model. Otherwise, the agreement is excellent.
The calculations have been made for a flexible bond chain with $r_0=6$ \AA\
and the root-mean-square bond length is around 12 \AA, a little longer
at low salt concentrations and a little shorter at high screening. As the
simulation results show the behaviour is almost identical to that of a
rigid bond chain with $a=12$ \AA. It is a general conclusion that
models with flexible, harmonic bonds give the same results as rigid bond
chains if the the actual, simulated root-mean-square monomer-monomer
separation is used to represent the bond length. A pair similar to
$r_0=6$ \AA/$a=12$ \AA\ is $r_0=9$ \AA/$a=24$ \AA.

When the Debye length is on the order of the bond length ($\kappa a \approx 1$)
the variational result crosses over to a third regime where $l_{p,e}$ is
proportional to $\kappa^{-2}$, i.e. we get a behaviour similar to the OSF
prediction, which applies to stiff chains. Although the chain is highly
flexible, it is stiff on the length scale of a bond and when $\kappa^{-1}$
is of this order, the intrinsic persistence length, $l_{p,0}=a$, dominates
the total persistence length. The OSF theory treats the electrostatic
interactions as a perturbation to an intrinsic bending energy,
which is quadratic in the bending angle.
Analysing the high
temperature expansion, Eq.~(\ref{ehT}), which is also a perturbational
calculation, we would get a $\kappa^{-2}$ dependence to the lowest order for the
entire range of salt concentrations if $l_{p,e}$ were plotted as a function of
$\sqrt{\xi_p}/\kappa$, but scaling with the bond length as in Fig.~\ref{fvar}
the expansion result curves upwards at lower salt concentrations.

The simulation results appear also to cross over into a third regime
at (ridiculously) high salt concentrations, although the exponent of $\kappa^{-1}$
seems to be closer to 1.5 and slowly increasing.
Seidel's large exponent, also 1.5, for $\kappa^{-1}$ may be due to the fact
that the Debye length
was smaller than the distance between neighbouring charges.\cite{Seidel:94}

Finally, the low temperature expansion, which
corresponds to a rigid rod, gives the expected almost constant persistence
length at low salt concentrations, but diverges when $\kappa R_{ee} \approx 1$
and the chain cannot bend as it is supposed to.
 
\subsection{Stiff polyelectrolyte}
Thus, we have shown that the OSF theory is not applicable for a flexible polyelectrolyte
where the electrostatic persistence length is of the same order as the total
persistence length. This should not come as a surprise, since the OSF theory
was derived for a chain with a significant intrinsic stiffness and where the
electrostatic persistence length only gives a minor contribution to the total
persistence length. 

\begin{figure}[t,b,p]
\hspace{2.5cm} \hbox{
        \psfig{figure=\PLOTDIR/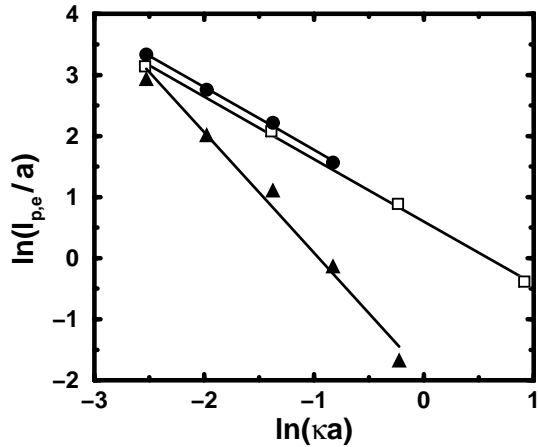,width=9.15cm}
}
\vspace{-2.5cm}
\caption{The electrostatic persistence length for a 320-mer with rigid bonds
         of $a=24$ \protect\AA\ and an angular potential with $l_{A}=\protect\sqrt{2000}$ \protect\AA\
         (filled circles) and $l_A=\protect\sqrt{20000}$ \protect\AA\ (filled triangles). The solid lines
         are least square fits, which have the slopes -1.0 and -2.0,
         respectively. Also shown is the corresponding chain without the
         angular potential (open squares).
}
\label{lpstiff}
\end{figure}

One way to introduce a non-electrostatic stiffness beyond the bond length is via 
an angular potential [see Eq. (\ref{eang})]. We have performed such simulations for 
a fully charged chain with 320 monomers and a fixed bond length $a=24$ \AA\
and found that the persistence length
still scales like $\kappa^{-1}$. With increasing $l_A$, however, the dependence changes
over to $\kappa^{-2}$ in agreement with the OSF result. Figure \ref{lpstiff}
shows how a chain with $l_A=\sqrt{2000}$ \AA\ and with an intrinsic persistence length,
$l_{p,0}=83$ \AA\ decays like $\kappa^{-1}$, just like the flexible chain
without the angular potential, i.e. $l_A=0$ \AA.
Upon increasing the stiffness to 
$l_A=\sqrt{20000}$ \AA\ and a $l_{p,0}=759$ \AA\ the $\kappa$ dependence shifts to
-2. The $\kappa$ dependence at high salt concentration still seems to be quadratic
although considerable numerical difficulties prevent an accurate analysis.

\section{CONCLUSIONS}
Simulating a flexible polyelectrolyte model that only has screened Coulomb
interactions, besides the bond interactions, we find a linear relationship
between the electrostatic persistence length and the Debye length in the
controversial region where the Debye length is shorter than the chain
dimensions ($\kappa^{-1} \ll R_{ee}$) and the interactions have become more
local. Furthermore, we also find that the electrostatic persistence length
is proportional to the square root of the strength parameter $\xi_p=\alpha^2
l_B/a$, which measures the internal interactions, i.e we have
$l_{p,e}\sim \sqrt{\xi_p}/\kappa$. This is in agreement with experiment,
but contradicts OSF theory, which predicts $l_{p,e}\approx\alpha^2 l_B/(2\kappa
a)^2$. 

The conclusions remain the same if different definitions of persistence length
are used, although $l_{p,x}$ [Eq.~(\ref{elpx})] may give erratic 
results when obtained by a simple fitting to the orientational correlation
function.

Variational calculations agree with simulations. A small difference is
seen at very high salt concentrations, where the variational results predict
a $\kappa^{-2}$ dependence as in the OSF theory. Here simulations give
an exponent $p$ between 1 and 2 for $l_{p,e} \sim \kappa^{-p}$.
In this region, $\kappa^{-1} \ll a$, and the electrostatic interactions can be
treated as a small perturbation to the harmonic interactions in the variational
ansatz, similar to the perturbation of the quadratic bending energy in the
OSF theory. A high temperature expansion carried to first order gives a
quadratic dependence throughout the concentration range.

Finally, for a stiff chain where the electrostatic persistence length only gives
a minor contribution to the the total stiffness, we find that theory due to Odijk
and Skolnick and Fixman is valid. Under these conditions the electrostatic persistence
length scales like $\kappa^{-2}$.


\bibliography{myref2}
\bibliographystyle{jcp2}

\end{document}